\documentclass[12pt]{article}

\usepackage[margin=1.25in]{geometry}
\usepackage[utf8]{inputenc}
\usepackage[T1]{fontenc}
\usepackage{hyperref}
\usepackage{url}
\usepackage{microtype}
\usepackage{setspace}
\usepackage{natbib}
\usepackage{titling}
\usepackage{authblk}

\pretitle{\begin{center}\large\bfseries}
\posttitle{\end{center}\vspace{-0.5em}}
\preauthor{\begin{center}}
\postauthor{\end{center}\vspace{-0.5em}}
\predate{\begin{center}\small}
\postdate{\end{center}}

\title{Position: The Pre/Post-Training Boundary Should Govern IP
in Industry--Academia ML Collaborations}

\author[1]{Dirk Bergemann}
\author[2]{Soheil Ghili}
\author[3]{Nitzan Mekel-Bobrov}

\affil[1]{Yale University \\ \texttt{dirk.bergemann@yale.edu}}
\affil[2]{Yale University \\ \texttt{soheil.ghili@yale.edu}}
\affil[3]{eBay Inc.\ \\ \texttt{nitzan@ebay.com}}

\date{Working Paper \\ May 2026}

\begin{document}

\maketitle
\thispagestyle{empty}

\begin{abstract}
Industry--academia ML collaborations routinely fail to launch---not for
scientific reasons, but because academics must publish while companies must
protect models trained on proprietary data, and no standard contract framework
resolves this tension. Because contracts are negotiated by legal departments
alone, many apparent legal disputes are incentive misalignment problems that only
scientists at the table can correctly diagnose. We propose PBOS
(Protect-the-Business / Open-Source-the-Science), a community-adoptable contract
template anchored to a single technically-grounded boundary: pre-training
artifacts (architectures, training code, benchmarks, untrained weights) are open
science; post-training artifacts (weights trained on proprietary data) are
business IP. This boundary is technically meaningful, legally clean, and
auditable---and could not have been drawn correctly without scientists at the
negotiating table. We argue the ML community should adopt PBOS as its default
contract for such collaborations.
\end{abstract}

\newpage
\setcounter{page}{1}

% -------------------------------------------------------

\section{Introduction}

The complementarity between industry and academia in machine learning is
obvious. Companies accumulate proprietary datasets---transaction logs, user
behavior, negotiation transcripts---at a scale and richness that no university
lab can replicate. Universities produce the algorithmic ideas, the graduate
talent, and the publication record that companies depend on to attract that
talent and advance the science. On paper, the incentives to collaborate are
overwhelming.

In practice, they are not enough. In the spring of 2024, a university research team set
out to build an AI negotiator that could learn to bargain on a major
peer-to-peer marketplace. The company had millions of real negotiation
transcripts; the researchers had the methods and the people to study them.
Together, they could advance both the science of strategic AI and the
company's product. Yet before the research could begin, the collaboration
was delayed for months in contract negotiations. Lawyers exchanged successive
document drafts across dozens of versions. The central question was not about
money, access, or scientific scope---it was about where to draw the line
between what could be published and what had to stay proprietary. That question
had no standard answer, because the ML community has never agreed on one.

This is not an isolated case. The friction between academic publication norms
and corporate IP protection is a structural feature of the landscape, not a
series of unlucky negotiations. Academics must publish to advance their
careers; companies must protect the models trained on their data to preserve
their competitive advantage. Neither side is acting irrationally. What is
missing is a shared contract framework that tells both sides where the line is
before the lawyers start drafting.

The absence of such a framework has a second, less obvious consequence: because
contracts are negotiated solely between legal departments, the technical
structure of the collaboration is invisible to the people writing the terms.
Many apparent legal disputes in this setting are not really legal problems---
they are incentive misalignment problems in disguise. Correctly diagnosing
them requires knowing what the scientists are building: which artifacts reveal
proprietary information and which do not, which outputs the company
legitimately needs to control and which the academic legitimately needs to
release. Lawyers alone cannot answer these questions. Scientists must be at the
table.\footnote{The collaboration described above produced the
contract framework studied in this paper precisely because researchers
participated directly in contract negotiation, proposing specific language
around what constitutes ``The Science'' (Section~\ref{sec:framework}).}

We propose PBOS (Protect-the-Business / Open-Source-the-Science) as a
community-standard contract template for ML collaborations involving
proprietary data, anchored to a single technically-grounded boundary:
pre-training artifacts---architectures, training code, benchmarks, and
untrained weights---are open science and must be releasable; post-training
artifacts---weights trained on proprietary data---are business IP and remain
under exclusive commercial control, with academics retaining only a narrow
research license to use them for evaluation and publication.\footnote{The
PBOS contract template, clause pack, and implementation guidance are available
at \url{https://github.com/soheil23/pbos-contract}. This paper is available
at \url{https://sites.google.com/view/soheil-ghili/pbos}.}

The pre/post-training boundary is technically meaningful (trained weights
encode proprietary data; untrained artifacts do not), legally clean (ownership
of each artifact class is unambiguous), and practically auditable (transfer logs and repository records make it
straightforward to verify what left corporate infrastructure). Crucially,
it is a boundary that
could not have been drawn correctly without ML researchers at the negotiating
table---it is a technical claim, not a legal one.

We further argue that the process of cross-disciplinary contract negotiation,
historically rare because scientists lacked fluency in legal language, is now
becoming scalable. AI tools allow researchers to engage with contract language
substantively without becoming lawyers. Scientific proposals need not be
legally perfect to be influential: even partial engagement shifts negotiations
toward incentive alignment and away from protracted contract revision cycles.

The rest of the paper is structured as follows. Section~\ref{sec:problem}
characterizes the structural incentive misalignment underlying collaboration
failure. Section~\ref{sec:framework} presents the PBOS framework in
detail. Section~\ref{sec:incentives} discusses why PBOS aligns incentives,
its community precedents, the boundary questions the framework must handle,
and why researcher participation in contract negotiation is a technical
necessity. Section~\ref{sec:conclusion} concludes with a call to action.

% -------------------------------------------------------

\section{The Problem: Structural Incentive Misalignment}
\label{sec:problem}

Industry--academia ML collaborations fail not because the parties disagree
about science, but because their institutional incentives are misaligned in
a predictable way. Each side has requirements that the other's institutional
constraints make difficult to accommodate.

\subsection{The Nature of the Incentive Mis-alignment}

Academic researchers require the ability to publish: to describe their
methodology, report their results, and release artifacts sufficient for
replication. Publication determines career advancement, grant eligibility,
and scientific reputation. A collaboration that ends without a publishable
paper, or that produces one only after stripping out the empirical
contributions, fails on the academic's core requirement.

Companies that provide proprietary data for model training require control
over the resulting artifacts. A model trained on proprietary data encodes
patterns derived from that data; if the trained weights leave corporate
infrastructure, or if a publication reveals sufficient detail to permit
reconstruction, the company's competitive position is weakened. Companies
therefore require exclusive control over any artifact that has absorbed
their data.

These two requirements are not inherently incompatible, but they conflict
in the absence of a shared framework for distinguishing which artifacts
belong to science and which belong to the business. Without such a
framework, each collaboration must resolve this question independently,
incurring the associated transaction costs each time.

\subsection{The Consequences of the Mis-Alignment}

The consequences of the coordination failure are not confined to the
transaction costs of individual negotiations. When a class of problems is
systematically inaccessible to academic researchers, the corresponding
research literatures do not form. Problems that require proprietary
behavioral data at scale---learning from real-world negotiations, modeling
large-scale user dynamics, training agents on production logs---remain
outside the academic research frontier, regardless of their scientific
importance. As industry develops products and methods in these domains,
the divergence between the applied frontier and the academic research
frontier widens over time.

The cost to companies is the foregone contribution of external research
capacity. Academic collaborators identify failure modes, propose
architectural improvements, and produce peer-reviewed results that internal
teams cannot generate independently. When collaboration fails to form,
companies bear this cost in addition to the direct costs of extended
contract negotiations.

The aggregate effect is a market failure in scientific
production~\citep{murray2007ip}: the transaction costs of contracting are
high enough to deter a class of collaborations whose social value
substantially exceeds those costs.

% -------------------------------------------------------

\section{PBOS: The Framework}
\label{sec:framework}

PBOS (Protect-the-Business / Open-Source-the-Science) is a contract template
for ML research collaborations involving proprietary data. It organizes
existing contractual instruments around a single technically-grounded
boundary---the pre/post-training distinction---that resolves the artifact
ownership question in advance. This section presents the technical
foundation of that boundary, the three contract provisions that
operationalize it, and its practical mechanics.

\subsection{The Pre/Post-Training Boundary}

PBOS resolves the artifact ownership question through a single technical
criterion: the pre/post-training boundary. Artifacts that have not been
trained on proprietary data---architectures, training pipelines, evaluation
benchmarks, loss functions---describe scientific methodology and contain no
information derived from the company's data. The same holds for models
trained on public or synthetic data: their weights reflect only the public
or synthetic distribution. Publication of these artifacts does not
compromise the company's competitive position.

Artifacts trained on proprietary data occupy a different informational
category. The weights of a trained model encode patterns derived from the
training distribution; depending on the architecture and procedure, they
may be susceptible to membership inference attacks~\citep{shokri2017membership},
reconstruction of training examples, or extraction of distributional
statistics the company regards as confidential.

The pre/post-training boundary has three properties that make it suitable
as a contractual criterion. First, it tracks the actual information content
of the artifact: the relevant question is whether proprietary data has
influenced the artifact's content, not the artifact's form or who produced
it. Second, it maps onto a legally clean ownership rule: untrained artifacts
are University Work Product; trained artifacts are the company's property.
Third, it is auditable: trained weights are large and remain on company
infrastructure by design, and transfer logs provide a straightforward record
of what left corporate servers.\footnote{In the collaboration described in
this paper, the distinction was made legally operative by specifying that
``The Science'' explicitly includes models that are ``untrained on data,''
``trained using only public data,'' or ``trained on data generated through
simulations using publicly available tools''---and that any model trained on
company data is by definition company Work Product, regardless of who wrote
the architecture.}

Where this boundary falls in a given pipeline---including multi-stage
settings such as fine-tuning or transfer learning---is a technical
determination that legal departments cannot make reliably without ML
expertise. This is addressed further in Section~\ref{sec:participation}.

\subsection{Operationalization in Contracts: The Three Pillars}

PBOS operationalizes the pre/post-training boundary through three contract
provisions.

\textbf{Pillar 1: Define the science explicitly.}
The contract must contain a definition of what counts as ``the science.''
This definition---not a vague reference to ``academic purposes'' but a
specific enumeration of artifact types---is the load-bearing element of
the whole framework. In the agreement underlying this paper, ``The Science'' was defined
as mathematical equations, model architectures, algorithmic descriptions,
and data-free visual exhibits, with an explicit carve-out for untrained
models and models trained on public or synthetic data. The definition must
be written with scientists in the room, because the line between a
``data-free'' and a ``data-derived'' artifact is a technical judgment that
legal language cannot make on its own.

\textbf{Pillar 2: Protect the business.}
Anything trained on proprietary data is the company's exclusive commercial
property. The university retains a narrow research license---sufficient to
run evaluations, generate results, and publish findings---but not to deploy,
distribute, or commercialize the trained model.\footnote{This asymmetry is
important and is often misunderstood. Academics \emph{need} trained weights
to do the science: to evaluate their methodology, measure performance, and
generate the empirical results that appear in a paper. They do not need to
own or release those weights. The research license covers use; the company
retains exclusive rights to distribution and commercialization.}
The company also receives a non-exclusive, royalty-free license to use
University Work Product---that is, the publicly released science-side
artifacts---for any purpose.

\textbf{Pillar 3: Open-source the science.}
The university commits to releasing pre-training artifacts---architectures,
training code, benchmarks, and the paper itself---under a permissive license.
This is not a concession; it is what academic researchers need to do anyway
to publish. Making it an explicit contractual commitment clarifies that the
company cannot subsequently object to the release of science-side artifacts
on confidentiality grounds. The science is, by definition, devoid of
proprietary information---and the contract says so.
The open-source commitment carries a further benefit on the company side:
a researcher who has publicly released the pre-training artifacts has no
remaining incentive to commercialize them. The collaboration produces a
scientific contributor, not a latent competitor. This removes a class of
risk that companies implicitly guard against in any research
partnership---and removes it structurally, through the terms of the
agreement, rather than through monitoring.

\subsection{How It Works in Practice}

The mechanics of a PBOS collaboration follow from the framework above.
Sensitive computation---training on proprietary data---takes place on company
infrastructure. Trained weights never leave corporate servers except under
the research license, which governs their use for evaluation only. When the
paper is ready, the university submits it to the company for a brief review
period (thirty days in the agreement underlying this paper) to verify that no
proprietary information has inadvertently entered a figure or table. The
company may request a short delay (up to ninety days) if the work contains
patentable subject matter; it may not suppress publication on the grounds
that ``The Science'' is confidential, because the contract already
established that it is not.

The contract is intentionally short. Brevity is not a stylistic choice; it
is a design criterion. A long contract creates ambiguity, because
every clause that is not a PBOS pillar is a potential source of ambiguity.
The goal is a document that both the engineers writing the code and the
lawyers reviewing it can read in an afternoon and agree on---because the
hard question (where is the line?) has already been answered by the
technical definition of ``The Science.''

% -------------------------------------------------------

\section{Discussion}
\label{sec:incentives}

PBOS resolves the incentive misalignment by codifying a bargain in which
each side retains what it requires and relinquishes only what it does not:
the company retains exclusive commercial rights to trained models; the
university retains full ownership of pre-training artifacts and a research
license sufficient for evaluation and publication. The subsections below
examine why researcher participation in contract negotiation is a
technical necessity, and situate the pre/post-training boundary as an
established norm in industry practice.

\subsection{The Necessity of Researcher Participation}
\label{sec:participation}

The load-bearing element of PBOS---Pillar 1, the explicit definition of
``The Science''---cannot be written correctly by legal departments alone.
Determining which artifacts encode proprietary data and which do not is a
technical determination: whether a published architecture reveals anything
about the training distribution, whether untrained weights constitute a
trade secret, and where the pre/post-training boundary falls in a given
pipeline are questions that require ML expertise to answer. In its
absence, contract language tends toward one of two defaults: broad
ownership assignments that encompass all research outputs including
pre-training artifacts, or vague language that defers the boundary
question to publication time.

The collaboration described in this paper illustrates the alternative.
The initial draft agreement granted the company ownership of ``all
results, reports, analytical models, inventions, software or
methodologies developed during the research''---formulation that would
have encompassed model architectures, training code, and any
publication, none of which encode proprietary
information.\footnote{%
The final agreement defined ``The Science'' as mathematical equations,
model architectures, algorithmic descriptions, and data-free visual
exhibits---explicitly including untrained models and models trained on
public data---and established that these materials are University Work
Product not subject to the company's ownership claim, regardless of
whether they were developed in the context of the collaboration.}
The ownership question was resolved when the researchers defined The
Science, applied the three pillars, and built the contract around that
definition. The contract itself was a standard agreement; what made it
work was the technical input that determined where to draw the line.

\subsection{The Pre/Post-Training Boundary is an Established Industry Norm, but Not in Cross-Institutional Contracts}

The distinction PBOS codifies in contract language is not new---it
already governs how technology companies manage their own ML researchers.
When DeepMind published the AlphaGo architecture in \textit{Nature} in
2016, the paper described the full methodology in reproducible
detail~\citep{silver2016alphago}; the trained weights were never released.
Google published the PaLM architecture paper in
full~\citep{chowdhery2023palm}; the trained model remained proprietary.
OpenAI published the GPT-3 paper~\citep{brown2020gpt3}; the weights were
never made public.
In each case, the company's researchers were permitted---indeed
encouraged---to disseminate the science, because the science alone
confers no competitive advantage. The competitive asset is the model
trained on the company's data and compute, and that asset stayed inside.

This is the pre/post-training boundary in operation, enforced implicitly
by employment agreements and internal policy rather than by explicit
contract. It works within firms precisely because it correctly identifies
what the company actually needs to protect. What PBOS contributes is
portability: it takes a norm that already governs intra-firm
researcher--company relations and writes it into a contract that works
across institutional boundaries, enabling the same bargain between a
university lab and a technology company that currently operates only
between a company and its own employees.

The boundary is robust to the fine-tuning case that might seem to
complicate it. When a collaboration begins from a public checkpoint and
fine-tunes on proprietary data, the base weights remain open science---
they encode no proprietary information---while the fine-tuned artifact,
which has absorbed patterns from the company's data, belongs to the
company. The criterion is not the origin of the weights but their
exposure to proprietary data. This formulation handles fine-tuning,
transfer learning, and multi-stage pipelines without case-by-case
adjudication.

% -------------------------------------------------------

\section{Conclusion}
\label{sec:conclusion}

The high rate of failure in industry--academia ML collaborations is not
an inevitable feature of the landscape. We have argued that it arises
from a specific and addressable cause: the absence of a shared contract
framework that resolves the ownership of research artifacts before
negotiations begin. The collaboration underlying this paper produced such a
framework---PBOS, anchored to the pre/post-training boundary---and this
paper is a proposal to generalize it.

The solution is a single technically-grounded criterion: the
pre/post-training boundary. Pre-training artifacts---architectures,
training code, benchmarks, untrained weights---contain no information
about the company's data and can be published freely. Post-training
artifacts---weights trained on proprietary data---encode patterns derived
from that data and must remain under the company's control. This boundary
is technically meaningful, legally clean, and already observed in
practice: technology companies routinely allow their own researchers to
publish architectures and methodology while keeping trained models
proprietary. What has been missing is a contract template that makes this
boundary explicit and portable across disparate institutions. PBOS is
that template, and it requires scientists at the negotiating table to
draw the line correctly---because the pre/post-training distinction is a
technical claim, not a legal one.

We suggest three directions for adoption.

\textit{For researchers:} when a collaboration is delayed in contract
negotiations, a technical document explaining which artifacts encode
proprietary information and where the boundary between publishable
science and protected IP lies will often move negotiations further than
additional contract revisions. PBOS provides the vocabulary and
conceptual framework for writing such a document.

\textit{For institutions:} technology transfer offices and legal
departments should consider adopting PBOS as a default starting template
for ML research collaborations involving proprietary data. Doing so does
not require abandoning institution-specific terms; it establishes a
known baseline that reduces the time and cost required to reach
agreement.

\textit{For the ML community:} the community has developed standard
frameworks for model documentation, dataset documentation, and
responsible licensing. Extending this norm to the contracts that govern
access to proprietary data for research purposes would reduce transaction
costs and broaden access to a class of scientifically important problems.
We offer PBOS as a starting point and invite challenge and refinement.

We close with a conjecture. The collaborations that high contracting
costs currently deter are not uniformly distributed across the AI
landscape. They are concentrated in domains where proprietary behavioral
data is essential: learning from real-world negotiations, modeling
large-scale user dynamics, training agents on production logs that no
academic lab could independently generate. If PBOS or a similar
framework is adopted, we expect the volume of industry--academia ML
collaborations to increase, with the largest gains in areas where AI
intersects with real human economic and social behavior---a part of the
scientific frontier that is currently underexplored relative to its
importance.

% -------------------------------------------------------

\bibliographystyle{abbrvnat}
\bibliography{references}

\end{document}